\documentclass[12pt]{article}
\usepackage{amssymb,amsmath,epsfig}

\begin{document}

\title{\bf Phantom Energy Accretion by Stringy Charged Black Hole
\footnote{Supported by the Higher Education Commission, Islamabad,
Pakistan through the Indigenous Ph.D. 5000 Fellowship Program
Batch-IV.\newline $^{**}$Email: msharif.math@pu.edu.pk}}

\author{M. Sharif$^{**}$, G. Abbas\\
Department of Mathematics, University of the Punjab,\\
Quaid-e-Azam Campus, Lahore-54590, Pakistan}
\date{}

\maketitle

We investigate the dynamical behavior of phantom energy near stringy
magnetically charged black hole. For this purpose, we derive
equations of motion for steady-state spherically symmetric flow of
phantom energy onto the stringy magnetically charged black hole. It
is found that phantom energy accreting onto black hole decreases its
mass. Further, the location of critical points of accretion is
explored, which yields mass to charge ratio. This ratio implies that
accretion process cannot transform a black hole into an extremal
black hole or a naked singularity, hence cosmic censorship
hypothesis remains valid here.

\

The astronomical observations of our universe provide the evidence
for the existence of unusual type of matter known as dark energy (DE
), which governs expansion of the universe.$^{[1-4]}$ It is
estimated that two-third of our universe is made up of DE, which has
large negative pressure and can drive the accelerated expansion of
the universe.$^{[5]}$ It exhibits some unusual properties such as
negative value of equation of state (EoS) parameter and violation of
energy conditions.$^{[6]}$ Numerous models are proposed as
candidates of DE such as vacuum energy, quintessence, phantom and
Chaplygin gas. The vacuum energy (cosmological constant) is the
simplest form of DE for which the EoS parameter is $\omega=-1$. The
quintessence and phantom are the forms of DE for which $\omega>-1$
and $\omega<-1 $, respectively.$^{[7-9]}$ Saridakis$^{[10]}$
discussed the theoretical limits on the EoS parameter in phantom
cosmology, and has also investigated the phantom evolution using
power-law potentials.$^{[11]}$ Leon and Saridakis $^{[12]}$ used the
phantom dark energy model to explore the acceleration and
coincidence problem in cosmology. Phantom energy violates the
dominant energy condition.

A massive object surrounded by a matter can capture particles of the
matter that passes within certain distance from the massive object.
This phenomena is termed as accretion of matter by the massive
objects. Bondi $^{[13]}$ originally formulated the problem of matter
accretion by the compact objects in Newtonian gravity. Michel
$^{[14]}$ is the pioneer who studied accretion of gas onto the
Schwarzschild black hole (BH) in the relativistic physics. Sun
$^{[15]}$ discussed phantom energy accretion onto BH in the cyclic
universe. Babichev {\it et al.}$^{[16]}$ have shown that BH mass
diminishes due to phantom accretion. Jamil {\it et al.}$^{[17]}$
have explored the effects of phantom energy accretion onto the
charged BH in 4D. They pointed out that if the mass of BH becomes
smaller (due to accretion of phantom energy) than its charge, then
cosmic censorship hypothesis is violated. The same conclusion was
deduced by Babichev {\it et al.}$^{[18]}$ in studying the phantom
accretion onto charged BH with generalized linear EoS and Chaplygin
gas EoS. In recent papers$^{[19,20]}$, we have studied phantom
accretion onto the Schwarzschild de-Sitter and 5D charged BHs.

The stringy BHs have been the subject of interest for the last few
years, due to the fact that string theory is clearly defined theory
of quantum gravity. General relativity with some new matter fields
as the dilaton and axion resembles to the low energy effective
theory. Kar$^{[21]}$ and Dasgupta {\it et al}.$^{[22]}$ studied
energy conditions and the kinematics of the geodesic flow for the
charged stringy BHs. Sharif$^{[23]}$  investigated the structure of
force and potential for the stringy BH by using the pseudo-Newtonian
formulation. Sharif and Waheed$^{[24]}$  discussed the re-scaling of
energy for this BH by using the approximate symmetry approach.
Radinschi and Ciobanu$^{[25]}$  explored the energy momentum
distribution for the charged BHs. Motivated by these studies, we
investigate phantom accretion by string magnetically charged BH.

In this study, we follow the formulation of Michel$^{[14]}$ to
discuss the phantom accretion. It is found that phantom accretion
cannot transform a BH into a naked singularity or an extremal BH, in
contrast to the Reissner-Nordstr$\ddot{o}$m (RN) BH. The
gravitational units (i.e., the gravitational constant $G=1$ and
speed of light in vacuum $c=1$) are used.

The stringy magnetically charged BH solution is given by$^{[26]}$
\begin{equation}\label{1}
ds^2=\frac{1-\frac{m}{r}}{1-\frac{Q^2}{mr}}dt^2
-\frac{1}{({1-\frac{m}{r}})({1-\frac{Q^2}{mr}})}dr^2
-r^2(d\theta^2+\sin^2{\theta}d\phi^2),
\end{equation}
where $m$ and $Q$ are the mass and charge of the BH. For $Q=0$ and
$m=2M$, we obtain the Schwarzschild solution. The black hole
horizons can be found by solving $g^{11}=0$ which leads to
\begin{eqnarray}\label{2}
r_{\pm}= \frac{(m^2+Q^2)\pm \sqrt{(m^2-Q^2)^2}}{2m},
\end{eqnarray}
where $r_\pm$ imply that there are two horizons $r_+=m$ and
$r_-=\frac{Q^2}{m}$ such that $r_+>r_-$. For $m^2=Q^2$, we obtain
$r_{+}=r_{-}\equiv m$, which is the case of an extremal charged BH.
Unlike RN BH for $m^2<Q^2$, both horizons exist and one cannot
obtain a naked singularity at $r=0$.

The energy-momentum tensor for perfect fluid reads
\begin{equation}\label{3}
{T_{{\mu}{\nu}}={({\rho}+p)}u_{\mu}u_{\nu}-pg_{\mu\nu}},
\end{equation}
where $\rho$ is the energy density, $p$ is the pressure and
$u^\mu=(u^t,u^r,0,0)$ is the four-vector velocity. We mention
here that $u^\mu$ satisfies the normalization condition, i.e.,
$u^\mu u_\mu =1$. The conservation of energy-momentum tensor
yields
\begin{equation}\label{4}
\frac{r^2u}{1-\frac{Q^2}{mr}}(\rho+p)\left(({1-\frac{Q^2}{mr}})({1-\frac{m}{r}})+u^2\right)^{1/2}=D_0,
\end{equation}
where $D_0$ is an integration constant and $u^{r}=u<0$ for inward
flow of phantom towards the BH .

The energy flux equation can be derived by projecting the
energy-momentum conservation law on the four-velocity, i.e., ${u_\mu
{\nabla}_\nu T^{\mu\nu}}$=0 for which Eq.(\ref{3}) leads to
\begin{equation}\label{5}
\frac{r^2u}{1-\frac{Q^2}{mr}}\exp(n)=-D_1,
\end{equation}
where $D_1>0$ is another integration constant which is related to
the energy flux and
$n=\int^{\rho_h}_{\rho_\infty}\frac{d\rho'}{\rho'+p(\rho')}$. Also,
$\rho_h$ and ${\rho_\infty}$ are densities of the phantom energy at
horizon and infinity. From Eqs.(\ref{4}) and (\ref{5}), we have
\begin{equation}\label{6}
(\rho+p)\left(({1-\frac{Q^2}{mr}})({1-\frac{m}{r}})+u^2\right)^{1/2}
\exp(-n)=D_2,
\end{equation}
where $D_2=-\frac{D_0}{D_1}=\rho_\infty +p(\rho_\infty)$.

The rate of change of BH mass due to phantom energy accretion
is$^{[18]}$
\begin{equation}\label{7}
\dot{M}=-4 \pi r^2 {T^r}_t.
\end{equation}
Using Eqs.(\ref{4}-\ref{6}) in the above equation, we obtain
\begin{equation}\label{8}
\dot{M}=4\pi D_1({\rho}_\infty +{p}_\infty),
\end{equation}
which implies that mass of BH decreases if $({\rho}_\infty
+{p}_\infty)<0$. Thus the accretion of phantom energy onto a BH
leads to decrease of the mass of BH. The phantom energy accretion
only diminishes mass and does not affect the charge of BH. That is
why RN BH is converted into a naked singularity by the phantom
accretion and CCH is violated. However, critical accretion process
mentioned below implies that CCH remains valid in this case. Since
all $p$ and $\rho$, violating dominant energy condition, must
satisfy this equation, it holds in general. We would like to mention
here that the above relation is same for all 4D spherically
symmetric BHs.

Here, we locate such points at which flow speed is equal to the
speed of sound during accretion. The fluid falls onto the BH with
monotonically increasing velocity along the particle trajectories.
We discuss the critical accretion. The conservation of mass flux,
${{\nabla}_\mu J^\mu}=0$, gives
\begin{equation}\label{9}
\frac{\rho u r^2}{1-\frac{Q^2}{mr}}=k,
\end{equation}
where $k$ is an integration constant. It is obvious that $k<0$ as
$u<0$ and all the other quantities are positive. Using Eqs.(\ref{4})
and (\ref{9}), we obtain
\begin{equation}\label{10}
\left(\frac{\rho +p}{\rho}\right)^2 \left(
({1-\frac{Q^2}{mr}})({1-\frac{m}{r}})+u^2\right)=k_1,
\end{equation}
where $k_1=(\frac{C_0}{k})^2$ is a positive constant.
Differentiating Eqs.(\ref{9}) and (\ref{10}) and eliminating
$d\rho$, we have
\begin{equation}\label{11}
\frac{dr}{r}\left(2V^2-\frac{\frac{1}{2}(\frac{M}{r}-\frac{2Q^2}{r^2}
+\frac{Q^2}{mr})}{({1-\frac{Q^2}{mr}})({1-\frac{m}{r}})+u^2}\right)+\frac{du}{u}
\left(V^2-\frac{u^2}{({1-\frac{Q^2}{mr}})({1-\frac{m}{r}})+u^2}\right)=0,
\end{equation}
where $V^2=\frac{dln(\rho+p)}{dln\rho}-1$. The critical points are
found by taking both the factors inside the brackets equal to zero.
Thus we obtain
\begin{eqnarray}\label{12}
&&{u_c}^2=\frac{1}{4m r_c^2}r_c(m^2+Q^2) - 2m{Q}^2,
 \\\label{12a}
&&{V_c}^2=\frac{r_c(m^2+Q^2) - 2m{Q}^2}{5mr^2-3r_c(Q^2+m^2)+2Q^2m}.
\end{eqnarray}
We find that the solutions of the above equations are obtained if
${u_c}^2>0$ and ${V_c}^2>0$ implying that
\begin{eqnarray}\label{14}
r_c(m^2+Q^2) - 2m{Q}^2>0, \quad 5mr^2-3r_c(Q^2+m^2)+2Q^2m>0.
\end{eqnarray}
The subscript $c$ is used to represent a quantity at a point where
speed of flow is equal to the speed of sound, such a point is called
a critical point. The second one in Eq.(\ref{14}) has the solution
\begin{equation}\label{16}
r_{c\pm}=3(m^2+Q^2)\pm\sqrt{(9m^4-22m^2Q^2+9Q^4)},
\end{equation}
which will be real if
\begin{equation}\label{16}
\frac{m^2}{Q^2}\geq\frac{1}{9}(11+2\sqrt{10})\approx 1.925.
\end{equation}
The location of the critical points near the BH can be determined by
the roots $r_{c\pm}$. For the solution about critical point, we
insert the value of $r_{c\pm}$ in the first one of Eq.(\ref{14}) and
obtain a unique inequality
\begin{equation}\label{17}
\frac{m^2}{Q^2}>\frac{1}{2}(3+\sqrt{5})\approx 2.618.
\end{equation}
Since $r_{c\pm}$ remain real for the above ratio, accretion is
possible through $r_{c\pm}$ as long as the above inequality is
satisfied.

In summary, there always exist two horizons for stringy magnetically
charged BH independent of the $m^2$  to  $Q^2$ ratio of BH. In other
words, whatever the mass to charge ratio would be, a stringy
magnetically charged BH cannot be converted into a naked
singularity. We would like to mention that for $\frac{Q^2}{m^2}>1$,
the horizons for the RN BH disappear but there does not exist such
ratio in the present case for which horizons disappear. The critical
accretion analysis implies that corresponding to two horizons there
exist two values of $r_{c\pm}$ which can play the role of critical
points if the mass and charge of BH satisfy
$\frac{m^2}{Q^2}>\frac{1}{2}(3+\sqrt{5})\approx 2.618$. It is
concluded that phantom accretion decreases the mass of BH and
converts it to an extremal BH if $m^2= Q^2$. However, during the
accretion, we have $\frac{m^2}{Q^2}>\frac{1}{2}(3+\sqrt{5})\approx
2.618$. Thus unlike RN BH, the stringy magnetically charged BH
cannot be transformed to an extremal charged BH or a naked
singularity and CCH remains valid here.

\end{document}